%% file: ERE2012-Kasner.tex
\documentclass[graybox]{svmult}

\usepackage{mathptmx}       
\usepackage{helvet}         
\usepackage{courier}        
\usepackage{type1cm}        
\usepackage{makeidx}         
\usepackage{graphicx}        
\usepackage{multicol}        
\usepackage[bottom]{footmisc}

\makeindex             

\begin{document}

\title*{Kasner Solution in Brans-Dicke Theory and its Corresponding Reduced Cosmology}
\author{S. M. M. Rasouli}
\institute{S. M. M. Rasouli \at Departamento de F\'{\i}sica,
Universidade da Beira Interior, 6200 Covilh\~{a}, Portugal,
\\\email{mrasouli@ubi.pt}}
%
%
\maketitle

\abstract{We present a brief review of the modified Brans-Dicke theory (MBDT) in arbitrary
dimensions, whereby the ($N+1$)-dimensional field equations reduce to the $N$-dimensional $(ND)$ configuration
with sources and an effective induced scalar potential.
We then investigate a generalized Bianchi type~I
anisotropic cosmology in $5D$ BD theory that leads to an extended Kasner solution. By employing the
original equations of MBDT, we probe the reduced Kasner cosmology on the hypersuface
with proceeding the investigations for a few cosmological quantities,
explaining their properties for some cosmological models.}

\section{Dimensional Reduction of Brans-Dicke Theory in Arbitrary Dimensions}
\label{sec:1}
The original motivation of the induced-matter
theory (IMT)~\cite{ponce-wesson-92}
was to achieve the unification of matter
and geometry. Recently,
the idea of the IMT has been employed for generalizing
BD theory, as a fundamental underlying theory, in
four~\cite{Ponce2-10}
and arbitrary dimensions~\cite{RFMinprogress}.
In the following, we present only a brief review of the latter.

The variation of the $(N+1)D$ BD action in vacuum with
respect to metric and BD scalar field, $\phi$, give the equations
\begin{equation}\label{(D+1)-equation-1}
G^{^{(N+1)}}_{ab}=\frac{\omega}{\phi^{2}}
\left[(\nabla_a\phi)(\nabla_b\phi)-\frac{1}{2}\gamma_{ab}(\nabla^c\phi)(\nabla_c\phi)\right]
+\frac{1}{\phi}\Big(\nabla_a\nabla_b\phi-\gamma_{ab}\nabla^2\phi\Big),
\end{equation}
\begin{equation}\label{(D+1)-equation-2}
\nabla^2\phi=0,
\end{equation}
where the Latin indices run
from $0$ to $N$; $\gamma_{ab}$ is the metric associated to the $(N+1)D$ space-time,
$\nabla^2\equiv\nabla_a\nabla^a$ and $\omega$ is a
dimensionless parameter.
Here, we have chosen $c=1$.

In the following, we only employ the equations of the MBDT
in arbitrary dimensions, which convey relations between the $(N+1)D$ field
equations to the corresponding ones with sources in $ND$ space-time in the
context of BD theory~\cite{RFMinprogress}.

We can find the reduced field equations on the $ND$
hypersurface by employing the BD field Eqs.~(\ref{(D+1)-equation-1}), (\ref{(D+1)-equation-2})
and a $(N+1)D$ space-time with a line element
\begin{equation}\label{global-metric}
dS^{2}=\gamma_{ab}(x^c)dx^{a}dx^{b}=
g_{\mu\nu}(x^\alpha,l)dx^{\mu}dx^{\nu}+
\epsilon\psi^2\left(x^\alpha,l\right)dl^{2}\,,
\end{equation}
where the Greek indices run from zero to $(N-1)$, $l$ is a
non-compact coordinate associated to $(N + 1)$th dimension,
the parameter $\epsilon=\pm1$ allows we to choose the
extra dimension to be either time-like or space-like, and $\psi$
is the another scalar field taken as a function of all the coordinates.

Here, we only present some of the reduced equations on
the $ND$ hypersurface. These equations will be described in
two separated parts with a short interpretation.
\begin{enumerate}
\item We
can construct the Einstein tensor on the hypersurface as
\begin{eqnarray}\label{BD-Eq-DD}
G_{\mu\nu}^{^{(N)}}&=&\frac{8\pi}{\phi}T_{\mu\nu}^{^{(\rm BD)}}+
\frac{\omega}{\phi^2}\left[({\cal D}_\mu\phi)({\cal D}_\nu\phi)-
\frac{1}{2}g_{\mu\nu}({\cal D}_\alpha\phi)({\cal D}^\alpha\phi)\right]\cr
&+&\frac{1}{\phi}\left[{\cal D}_\mu{\cal D}_\nu\phi-
g_{\mu\nu}{\cal D}^2\phi\right]-g_{\mu\nu}\frac{V(\phi)}{2\phi},
\end{eqnarray}
where ${\cal D}_\alpha$ is the covariant derivative on $ND$ hypersurface,
which is calculated with $g_{\alpha\beta}$ , and ${\cal D}^2\equiv{\cal D}^\alpha{\cal D}_\alpha$.

The above equations correspond to the BD equations, obtained from
the standard BD action containing a scalar potential, but here
there are some differences which we clarify them in the following
\begin{itemize}
\item
The quantity introduced by $V(\phi)$ is actually the effective induced scalar potential
on the hypersurface which will be determined by a relation in part 2.
\item
The quantity $T_{\mu\nu}^{^{(\rm BD)}}$, can be interpreted as
an induced energy-momentum tensor (EMT)
for a BD theory in $N$-dimensions and it, in turn, contains three components as
\begin{eqnarray}\label{matt.def}
T_{\mu\nu}^{^{(\rm BD)}}\equiv T_{\mu\nu}^{^{(\rm I)}}+T_{\mu\nu}^{^{(\rm \phi)}}
+\frac{1}{16\pi}g_{\mu\nu}V(\phi),
\end{eqnarray}
where $8\pi/\phi T_{\mu\nu}^{^{(\rm I)}}$ is the same as the induced EMT appearing in IMT, while
\begin{eqnarray}\label{T-phi}
\frac{8\pi}{\phi}T_{\mu\nu}^{^{(\rm \phi)}}\equiv
\frac{\epsilon\phi_{_{,N}}}{2\psi^2\phi}\left[g_{\mu\nu,}{}_{_{N}}
+g_{\mu\nu}\left(\frac{\omega\phi_{_{,N}}}{\phi}-g^{\alpha\beta}g_{\alpha\beta,}
{}_{_{N}}\right)\right],
\end{eqnarray}
where $A{_{,N}}$ is the partial derivative of the quantity $A$ with respect to
$l$.
\end{itemize}
\item
The wave equation on the hypersurface is given by
\begin{eqnarray}\label{D2-phi}
{\cal D}^2\phi=\frac{8\pi T^{^{(\rm BD)}}}{(N-2)\omega+(N-1)}+
\frac{1}{(N-2)\omega+(N-1)}\left[\phi\frac{dV(\phi)}{d\phi}-\frac{N}{2}V(\phi)\right],
\end{eqnarray}
where
\begin{eqnarray}\label{v-def}
&\phi&\frac{dV(\phi)}{d\phi}\equiv-(N-2)(\omega+1)
\left[\frac{({\cal D}_\alpha\psi)({\cal D}^\alpha\phi)}{\psi}
+\frac{\epsilon}{\psi^2}\left(\phi_{_{,NN}}-
\frac{\psi_{_{,N}}\phi_{_{,N}}}{\psi}\right)\right]\\\nonumber
&-&\!\!\!\frac{(N-2)\epsilon\omega\phi_{_{,N}}}{2\psi^2}
\left[\frac{\phi_{_{,N}}}{\phi}+g^{\mu\nu}g_{\mu\nu,}{}_{_{N}}\right]
+\frac{(N-2)\epsilon\phi}{8\psi^2}
\left[g^{\alpha\beta}{}_{_{,N}}g_{\alpha\beta,}{}_{_{N}}
+(g^{\alpha\beta}g_{\alpha\beta,}{}_{_{N}})^2\right].
\end{eqnarray}

\end{enumerate}

 Actually in this approach, the $(N+1)D$ field equations
(\ref{(D+1)-equation-1}) and (\ref{(D+1)-equation-2})
split naturally into four sets of equations on every $ND$ hypersurface,
in which we only have introduced the two sets (\ref{BD-Eq-DD}) and (\ref{D2-phi}).
Regarding the geometrical
interpretation of the other two sets, we will not discuss
them and leave them for next paper in this series.

In the following, we investigate the Kasner solution in BD theory in a $5D$ space-time;
then, as an application of the MBDT in cosmology, we present
the properties of a the reduced cosmology on the hypersurface.

\section{Kasner Solution in Brans-Dicke Theory and its Corresponding Reduced Cosmology}
\label{K. Solution}

We start with the generalized Bianchi type~I anisotropic model
in a $5D$ space-time as
\begin{equation}\label{bulk-metric}
dS^{2}=-dt^{2}+\sum^{3}_{i=1}a_i^{2}(t, l)dx_i^{2}+h^{2}(t, l)dl^{2}\,,
\end{equation}
where $t$ is the cosmic time, $(x_1, x_2, x_3)$ are the Cartesian
coordinates, $l$ is the non-compactified extra dimension, and $a_i(t, l)$, $h(t, l)$ are
different cosmological scale factors in each of the four
directions. We assume that there is no matter in $5D$
space-time and $\phi=\phi(t, l)$. In addition,
based on the usual spatial homogeneity, we solve the field
equations (\ref{(D+1)-equation-1}) and (\ref{(D+1)-equation-2})
by assuming separation of
variables as
\begin{equation}\label{sep.eq}
\phi(t,l)=\phi_0 t^{p_{0}}l^{s_{0}},\hspace{6mm}
a_i(t,l)\propto t^{p_{i}}l^{s_{i}},\hspace{6mm} h(t,l)=h_o
t^{p_{4}}l^{s_{4}},\hspace{6mm}
\end{equation}
where $h_0$ and $\phi_0$ are constants, and the $p_{_a}$'s and
$s_{_a}$'s ($a=0,2,3,4$ ) are parameters satisfying field equations. By replacing the {\it
ansatz} (\ref{sep.eq}) into Eq.~(\ref{(D+1)-equation-2}) and using~(\ref{bulk-metric}), we get
five classes of solutions.
In the following, we are interested to investigate just
the solutions that leads to a generalized Kasner relations in five
dimensions. Also, we then apply the MBDT to obtain the corresponding
reduced cosmology on a $4D$ hypersurface.

In order to have consistency and ignoring the trivial solutions we set $p_4\neq 1$ and
$s_4\neq -1$. Also for simplicity, we assume $h_o=1$ in {\it ansatz} (\ref{sep.eq}).
Hence after a little manipulation, we can obtain the following relations among the
generalized Kasner parameters~\cite{RFS11}
\begin{eqnarray}\label{kas-rel1}
\sum^{4}_{a=0}p_{_a}&=&1, \hspace{5mm}
(\omega+1)p_{0}^{2}+\sum^{4}_{m=1}p_m^{2}=1,\hspace{5mm}
\sum^{3}_{\mu=0}s_{\mu}=1+s_{4},\\\nonumber
(\omega+1)s_{0}^{2}&+&\sum^{3}_{i=1}s_i^{2}=(1+s_{4})^{2},\hspace{5mm}
(\omega+1)p_{0}s_{0}+\sum^{3}_{i=1}p_is_i=p_{4}(1+s_{4})\,.
\end{eqnarray}
Eqs.~(\ref{kas-rel1}) lead to a few constrains, so that there are only five
independent relations among the
Kasner parameters, designated as the generalized
Kasner relations in $5D$ BD theory.

In what follows, for the sake of brevity,
we would like to present a brief review of the
results of the reduced Kasner cosmology on the $4D$ hypersurface~\cite{RFS11}:
the pressure and energy density of the specified induced matter on any $4D$
hypersurface can be derived from (\ref{matt.def}).
These results show that, in general, we cannot consider it as a perfect fluid.
By applying (\ref{v-def}), (\ref{sep.eq}) and (\ref{kas-rel1}), the induced
scalar potential is obtained to be either in the power law or in the logarithmic form, in which we only
investigate the properties of the former.
 The properties of a few cosmological quantities as well as
physical quantities such as the average scale factor, the mean
Hubble parameter, the expansion scalar, the shear scalar and the
deceleration parameter have been studied. First,
these quantities have been derived in terms of the
generalized Kasner parameters. Then, we find that
the induced EMT satisfies the barotropic equation of state, where the equation
of state parameter, $w$, is a function
of the Kasner parameters. And thus, the evolution of all the
quantities has been represented with respect to $w$, $\omega$ and the deceleration parameter, $q$.
We then probe the quantities, in the general case, versus $q$, $t$ and $\omega$ for the stiff fluid and the
radiation-dominated universe. We have shown that, for both of the fluids, there is an expanding
universe commenced with a big bang, and there is a horizon for each of them. Also, we have shown that the rate of
expansion slows down by time. By employing the weak energy condition, the allowed (or the well-behaved)
ranges of the deceleration and the BD coupling parameters have been obtained for each of
the fluids. The behavior of the quantities,
in the very early universe and the very large time show that the models yield
empty universes when the cosmic time tends to infinity.
However, both of the models, in general, do not approach isotropy for large values
of the cosmic time.
\begin{acknowledgement}
I sincerely appreciate to Prof. Paulo Vargas Moniz for a critical reading of the letter.
I am supported by the Portuguese Agency Funda\c{c}\~{a}o para a
Ci\^{e}ncia e Tecnologia through the fellowship SFRH/BPD/82479/2011.
\end{acknowledgement}

\input{referenc}

\end{document}

%% file: referenc.tex
%
%